\documentclass[11pt,twoside]{article}
\usepackage{asp2004}
\usepackage{psfig}
\usepackage{epsf}
\usepackage{graphics}
\usepackage{lscape}
\begin{document}
\title{GMC Formation in M33: the Role of Pressure}
\author{Leo Blitz and Erik Rosolowsky}
\affil{Radio Astronomy Laboratory, University of California, Berkeley,
94720}
\begin{abstract}

We present a complete map of the Giant Molecular Clouds in M33 and
show that they lie exclusively on filaments of HI.  The GMCs are 
localized on the extended HI filaments, and HI filaments extend
well beyond the radius where the GMCs are found, suggesting that the
H$_2$ forms from the HI and not vice-versa.  We propose that the fraction of
neutral gas that is molecular at a particular radius is determined
primarily by the ambient hydrostatic pressure at that radius.  This
leads to a prediction that the transition radius, the radius where the
mean surface density of HI and H$_2$ are equal, occurs at a constant
value of the stellar surface density.  We show that for a sample of 26
galaxies, the stellar surface density at the transition radius is
constant to 40\%.

\end{abstract}
\thispagestyle{plain}
\section{Introduction}

The star formation history of galaxies is determined by how Giant
Molecular Clouds form.  Stars are known to form only in molecular
clouds; it is only in molecular clouds that gas temperatures are low
enough and gas densities are high enough that the Jeans mass is in the
range of individual stellar masses.  Nevertheless, GMC formation is
poorly understood.  A number of different formation mechanisms
have been proposed, but there is no agreement on which mechanism
dominates. It is not yet possible, for example, to predict which
galaxies will be molecule rich, and what the distribution of
molecular gas will be.  This shortcoming makes the goal of
understanding the star formation history of the Universe elusive at
the present time.  

One approach in narrowing the range of possible formation mechanisms
is to observe individual molecular clouds in other galaxies. By
determining molecular cloud properties over a large range of
ambient conditions, both within a given galaxy and from
galay-to-galaxy, it may be possible to see how the cloud properties
change with the physical environment. Ideally, one would like to have
a complete catalogue of GMCs for an entire galaxy. With present-day
instruments, this is possible only within the Local Group and requires
large blocks of observing time.    

So far, only five galaxies have been surveyed with high enough
resolution to catalogue individual GMCs.  In the Milky
Way, although the entire disk has been surveyed for CO, it has
not been possible to get a complete catalogue of individual GMCs
because CO line blending in particular directions (e.g. the tangent
points) makes it impossible to untangle individual GMCs from one
another.  M31 (e.g. Muller \& Gu{\' e}lin 2003) presents a similar
problem because of its high inclination, though the problem is not quite
as severe as it is in the Milky Way. Fukui et al. (2001) succeeded in
making a complete survey of GMCs in the LMC, the first such survey in
any galaxy.  Engargiola et al. (2003; hereafter EPRB03) have made a
complete survey of GMCs in M33, the first such survey in a spiral
galaxy;  both surveys have comparable linear resolution.  A complete
survey of GMCs has also been made in IC 10, but is as yet
unpublished (Leroy et al., in preparation).  Complete, published GMC
catalogs are thus available for only two galaxies.   In this
paper, we describe some of the results of the EPRB03 survey of M33 and
examine the implications for GMC formation.

\section{The M33 Map}

\begin{figure}[!ht]
\plotone {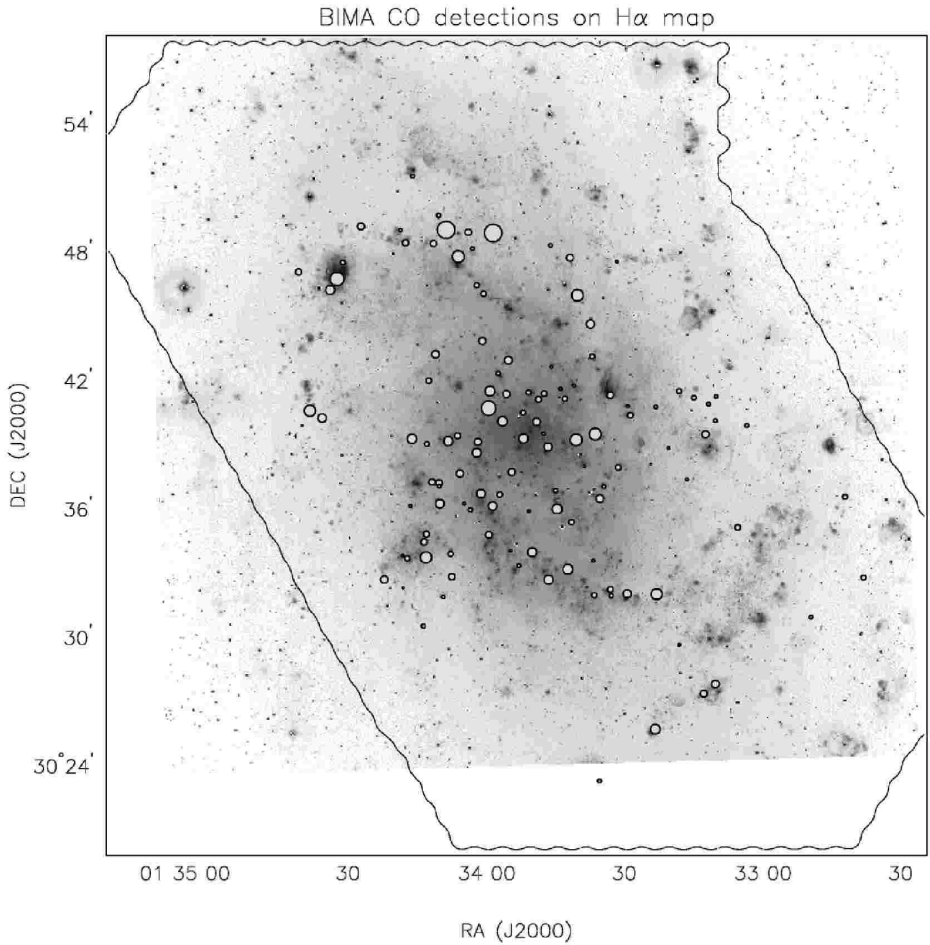}
\caption{GMCs in M33 plotted with white circles on the H$\alpha$ image of 
Cheng et al. (1996).  The area of each circle is proportional to the 
molecular mass. The irregular line shows the limits of the surveyed
region.}
\end{figure}

Figure 1 shows an H$\alpha$ image of M33 by Cheng et al. (1996).  Also
shown are the GMCs detected in the J = 1-0 transition of CO with the
BIMA Array by EPRB03. The CO map is a mosaic of 759 pointing centers
and covers about 0.25 deg$^2$.  The angular resolution of each
pointing center is 13\arcsec, which corresponds to a linear resolution
of 54 pc at the 850 kpc distance of M33.  A resolution of at least
$\sim$50 pc is desirable for GMC surveys because it minimizes
source confusion within the beam and makes direct comparison
between galaxies relatively straightforward:  50 pc is roughly the
mean diameter of GMCs in the Milky Way (e.g. Blitz 1993).  The total
integration time was about 15 minutes per pointing center, leading to
an rms noise of less than 0.5 Jy beam$^{-1}$ in the central half of
the mosaic.  Although some clouds are detected with masses as low as 4
$\times 10^4~\rm {M}_{\sun}$, the catalogue of GMCs is complete to a
mass of 1.5 $\times 10^5~\rm {M}_{\sun}$.

We find a steep mass function: $dN/dM \propto {M}^{-2.6}$,
significantly different from the mass function found in the Milky Way,
which has a power law index of -1.6 (e.g. Blitz 1993).  To highlight
this difference we note that the Milky Way is
estimated to have more than 100 GMCs with $M >$ 1.8 $\times 10^6~{\rm
M}_{\sun}$ (Dame et al. 1986); the highest mass GMC in M33 has a mass of only 7
$\times 10^5~{\rm M}_{\sun}$.  M33 has about 20\% of the mass of the
Milky Way, and about 5\% of its H$_2$ mass.  If the M33 mass function 
were the same as that in the Milky Way, we would expect 9 GMCs
with a mass greater than 7 $\times 10^5~{\rm M}_{\sun}$ scaling by the
relative  H$_2$ mass; none is observed.  The
difference in the mass functions between the two galaxies thus appears to be
quite significant. Nevertheless, the properties of the individual GMCs
appear to be quite similar in the two galaxies (Rosolowsky et al.
2003).

Unlike the Milky Way, the steep mass function for M33 suggests that
most of the molecular mass is at the low mass end of the spectrum.
However, a finite molecular mass for M33 requires that there be a
``turnover" mass where the power law index of the mass function becomes
shallower than -2.  EPRB03 show that this mass is 4--7 $\times
10^4~\rm{M}_{\sun}$.  This turnover mass is where most of the mass in
GMCs in M33 resides, and can be considered to be a characteristic mass
of the GMCs in this galaxy.

\section{Comparison with HI}

\begin{figure}[!ht]
\plotone{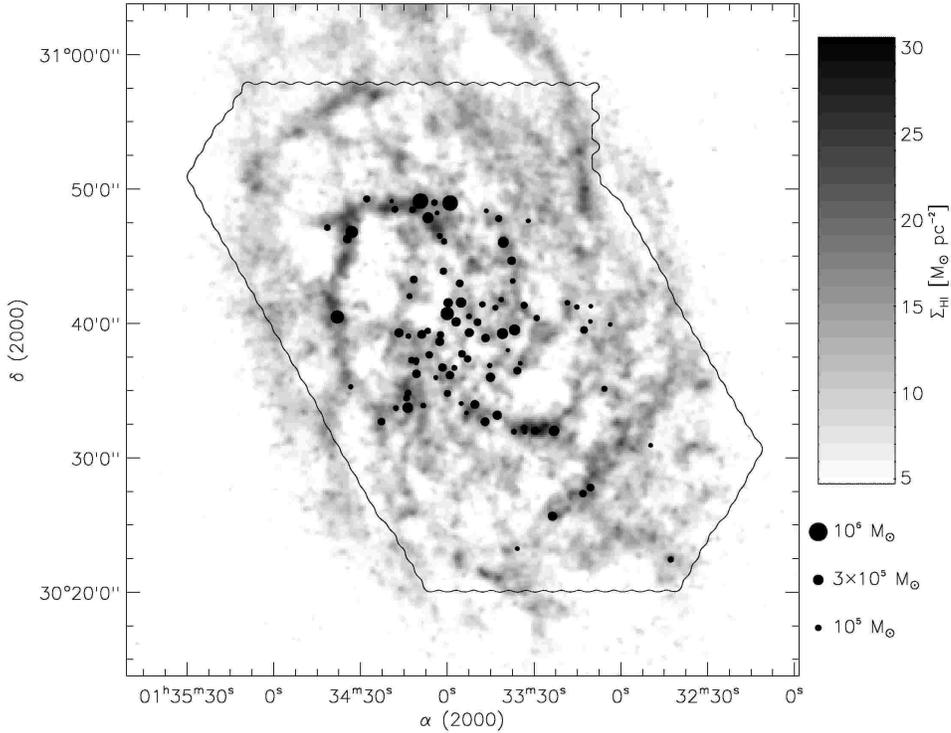}
\caption{Image of HI 21 cm emission from Deul \& van der Hulst (1987) 
with the GMCs overlaid.  Note that all GMCs lie on filaments of atomic
gas. As in Figure 1, the thin irregular line indicates the survey
boundary.}
\end{figure}

Figure 2 is a map of the GMCs in M33 overlaid on the HI map of Deul \&
van der Hulst (1987).  The HI is largely arranged in long, stringy filaments.
Remarkably, the GMCs are found only on the filaments.  Because the
filaments extend far beyond the radius where the GMCs are detected and
because the CO is far more localized than the filaments, any causal
relationship between the two gas phases implies that the
molecular gas formed from the HI.    

The close association between the GMCs and the HI filaments can be
used to set an upper limit on the GMC lifetimes.  The rms difference
between the CO and HI velocity centroids is 8 km s$^{-1}$ and the
typical width of an HI filament is $\sim$200 pc.  If the cloud
lifetimes were significantly longer than 20 Myr, the clouds would
drift off the filaments, destroying the good observed correlation.

The GMCs are closely associated with the large,
bright HII regions as can be seen in Figure 1.  The GMCs shown in Figure 1
contribute 36\% of the total flux in the galaxy.  These clouds are
contiguous with HII regions that contribute 36\% of the total
H$\alpha$ flux (EPRB03).  Thus the two distributions are consistent
with essentially all of the emission from HII regions being associated with GMCs.

The larger filaments often look as if they form the boundaries of
holes, raising the question: are the filaments actually the edges of
shells produced by energetic activity at their centers?  The ``holes"
in M33 have been catalogued by Deul \& den Hartog (1990) who find that
small holes $<$ 200 pc in diameter are statistically correlated with
OB associations. These may have been evacuated by the effects of
stellar winds and supernova explosions.  The large holes evident in
Figure 2, have diameters of 500 -- 1000 pc, and require energies of
$\ga 10^{53}$ ergs. Their lifetimes are $\la$ 3 $\times 10^7$ y
because they have not yet been sheared out by the differential
rotation of the galaxy.  If these holes have been evacuated by the
effects of star formation, there would be bright remnant stellar
clusters at the center of each feature, and these are not seen.  The
filaments therefore do not appear to have been produced by central
explosions or winds, but are generated by some mechanism
related either to spiral structure formation, or to some other gas dynamical
instability.

At all radii the azimuthally averaged atomic surface gas density is
greater than that of the H$_2$ surface density; near the center, the
ratio is about 5 and increases with increasing radius.  That is, the
surface density of HI is nearly constant with radius,  but that of the
CO decreases exponentially from the center with a scale length
comparable to the stellar scale length (EPRB03).  The HI extends to at
least twice the radius where detectable molecular clouds are found. 

It thus appears as if the filaments form first and they gather enough
gas together in one place to form the star-forming GMCs.  But there is
little radial variation in the surface density of HI. It is apparent
in Figure 1 that although the surface density in the HI filaments in
the outer half of M33 is similar to that in the inner part of the
galaxy, the outer galaxy filaments are nearly devoid of GMCs.  Furthermore,
the surface density along the filaments and from filament to filament
anywhere in the galaxy is also fairly constant.  Why then does the
molecular gas in the galaxy show such a strong radial dependence?

\section{The Role of Pressure}

Several authors have suggested that hydrostatic gas pressure
determines the molecular fraction at a given radius in a galaxy.
Spergel \& Blitz (1992), for example, argued that the extraordinarily
large molecular gas fraction at the center of the Milky Way is
plausibly the result of the very high hydrostatic pressure in the
Galactic bulge.  Elmegreen (1993) suggested on theoretical grounds
that the ratio of atomic to molecular gas in galactic disks results
from both the ambient hydrostatic pressure as well as the mean
radiation field.  The dependence on the pressure is steeper than that
of the radiation density ($f_{mol} \propto P^{2.2} j^{-1.1}$), and
ought to be the dominating factor.  Observationally, Wong and Blitz
(2002) showed that the radial dependence of the atomic to molecular
gas ratio in seven molecule rich galactic disks can be understood to
be the result of the variation of interstellar hydrostatic pressure
(with $f_{mol} \propto P^{0.8}$).

Using this as an operating hypothesis, let us assume that
$N(H_2)/N(HI)$, the ratio of H$_2$ surface density to HI surface
density, is determined {\it only} by hydrostatic pressure, $P_{ext}$.
In a thin disk with isothermal stellar and gas layers, and where the
gas scale height is much less than the stellar scale height as is
typical in disk galaxies, 

\begin{equation}
P_{ext} = (2G)^{0.5}\Sigma_g v_g\{{\rho_*}^{0.5} + (\frac{\pi}{4} 
\rho_g)^{0.5}\}     
\label{fullpressure}
\end{equation}

\noindent where $\Sigma_{g}$ is the total surface density of the gas,
$v_g$ is the velocity dispersion of the gas, $\rho_*$ is the midplane
suface density of stars, and $\rho_g$ is the midplane
suface density of gas.  In
most galaxy disks, $\rho_*$ is much larger than $\rho_g$, except in
the far outer parts of a galaxy where the stars become quite rare.  In
the solar vicinity, for example, $\rho_*$ = 0.1 M$_{\sun}$ pc$^{-3}$,
but $\rho_g \simeq$ 0.02 M$_{\sun}$ pc$^{-3}$ (e.g. Dame 1993).
For a self-gravitating stellar disk, $\Sigma_* = 2\sqrt2 \rho_* h_*$, 
where $h_*$ is the stellar scale height.  Thus, neglecting $\rho_g$,
Equation 1 becomes: 

\begin{equation}
P_{ext} =  0.84 (G \Sigma_*)^{0.5}\Sigma_g \frac {v_g} {(h_*)^{0.5}} \
\label{approxpressure}
\end{equation}

There is ample evidence that both $v_g$ and $h_*$ are both constant
with radius in 
disk galaxies.  Evidence that $h_*$ is
approximately constant comes from observations of edge-on
galaxies by van der Kruit and Searle (1981a,b; 1982a,b).  Measurements of the 
HI velocity dispersion in face-on galaxies indicate that $v_g$ is
approximately constant with radius (van der Kruit and Shostak
1982, 1984; Shostak and van der Kruit 1984).

By assumption,
\begin{equation}
N(H_2)/N(HI) = f(P_{ext}) \
\end{equation}
\noindent Thus, since $(v_g/\sqrt {h_*})$ is approximately constant within a
galaxy, then 
\begin{equation}
N(H_2)/N(HI) = f(\Sigma_g\Sigma_*)
\end{equation}

Wong and Blitz (2002) show that for a number of gas-rich spiral
galaxies, the N(HI) is nearly constant and saturates at a value of
about 6 $\times 10^{21}$ cm$^{-2}$, or about 3 magnitude of visual
extinction, as is true for M33.  Thus, the location in a galaxy where
$N(H_2)/N(HI)$ is unity, what we call the transition radius, occurs at
a fixed value of $\Sigma_g$.  This, in turn, implies that the
transition radius depends only on $\Sigma_*$. If $P_{ext}$ determines
$N(H_2)/N(HI)$, then $\Sigma_*$  should be constant
for all galaxies.

We investigate this prediction by determing the transition radius in
19 galaxies for which there is sufficient data from both CO and HI
maps to determine the location of the transition radius.  We then use
the 2MASS data to determine the $K$-band surface density at the
transition radius, and convert to stellar surface density using a
mass-to-light ratio of 0.5 M$_{\sun}/L_{\sun}$.  The
results are shown in Figure 3.

\begin{figure}[!ht]
\plotfiddle {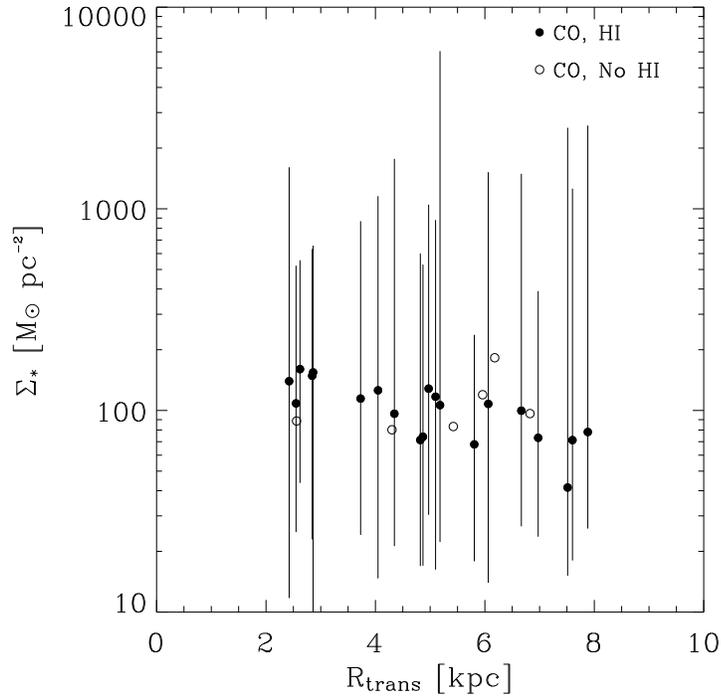}{4.0in}{0}{50}{50}{-175}{-30}
\caption{Plot of the location of the transition radius of a galaxy as
a function of stellar surface density at the transition radius.  The
thin vertical lines gives the range of stellar surface density in the
galaxy.  The open circles are 6 galaxies for which good CO maps exist
but not good HI maps.  In these latter galaxies, it is assumed that
the N(HI) is constant at a value of 2 $\times 10^{21}$ cm$^{-2}$. }
\end{figure}

Figure 3 shows a remarkable constancy of $\Sigma_*$, with a mean value
of 100 $\pm$ 40 M$_{\sun}$ {pc}$^{-2}$. This constancy is particularly
noteworthy because the range of stellar surface density is about
100 for these galaxies, and the range of transition radii varies by
a factor of 10.   

As a check, one can calculate the value of $\Sigma_*$ at the
transition radius in the Milky Way, scaling the measured $\Sigma_*$ at
the distance of the Sun (35 M$_{\sun}$ pc$^{-2}$ - Binney \& Merrifield
1998), and a radial scale length for the stars of 3 kpc (Spergel,
Malhotra \& Blitz 1996; Dehnen \& Binney 1998).  The
transition radius for the Milky Way occurs at a galactocentric distance
of about 4 kpc (Dame et al. 1993), or about 1.3 scale lengths inward
of the Sun.  This converts to a stellar surface density of 132
M$_{\sun}$
pc$^{-2}$, in good agreement with the measurements in other galaxies.


These arguments suggest that hydrostatic pressure may be the dominant
factor in determining how much atomic gas is converted to molecular
gas, on average, at a given radius.  Taken together with the gas
distribution in M33, the following picture emerges.  First, various
hydrodynamic and gravitational processes collect the gas in a disk galaxy
into filaments.  Some fraction of the gas gathered together
into filaments is turned into molecular gas. How much depends on radius
because of the monotonic, large-scale decrease in hydrostatic pressure
with radius.  In a galaxy like M33, with a low central surface
density, the gas may remain predominantly atomic throughout.  However,
in galaxies with prominent bulges and with high central stellar
surface densities, an overwhelmingly large fraction of the gas may
become molecular.  This may explain, for example, why the
molecular-to-atomic gas ratio in the center of the Milky Way is as
high as $\sim 10^{3}$, and why other bright galaxies have such large,
extended regions where the molecular gas is the dominant phase of the
ISM.

If hydrostatic pressure is the primary factor in determing the
surface density of molecular gas within galaxies, it brings us closer,
but not all the way to understanding how GMCs form.  One still needs to
understand why the gas forms into individual clouds, why they have
the masses that they do, and why their mass functions are different in
different galaxies.  
 


\end {document}